\documentclass[twocolumn]{aastex631}

\usepackage{amssymb,mathtools}

\shorttitle{Space Weather-driven Variations in Ly$\alpha$ Absorption in AU Mic b}
\shortauthors{Cohen et al.}

\graphicspath{{./}{figures/}}

\begin{document}

\title{Space Weather-driven Variations in Ly$\alpha$ Absorption Signatures of Exoplanet Atmospheric Escape:  MHD Simulations and the Case of AU Mic b}

\correspondingauthor{Ofer Cohen}
\email{ofer\_cohen@uml.edu}

\author[0000-0003-3721-0215]{Ofer Cohen}
\affiliation{Lowell Center for Space Science and Technology, University of Massachusetts Lowell, 600 Suffolk Street, Lowell, MA 01854, USA}

\author[0000-0001-5052-3473]{Juli\'{a}n D. Alvarado-G\'{o}mez}
\affiliation{Leibniz Institute for Astrophysics Potsdam, An der Sternwarte 16, 14482 Potsdam, Germany}

\author[0000-0002-0210-2276]{Jeremy J. Drake}
\affiliation{Center for Astrophysics \text{\textbar} Harvard \& Smithsonian, 60 Garden Street, Cambridge, MA 02138, USA}

\author[0000-0001-7944-0292]{Laura M. Harbach}
\affiliation{Astrophysics Group, Department of Physics, Imperial College London, Prince Consort Rd, London, SW7 2AZ, UK}

\author[0000-0002-8791-6286]{Cecilia Garraffo}
\affiliation{Center for Astrophysics \text{\textbar} Harvard \& Smithsonian, 60 Garden Street, Cambridge, MA 02138, USA}

\author[0000-0002-5456-4771]{Federico Fraschetti}
\affiliation{Dept. of Planetary Sciences-Lunar and Planetary Laboratory, University of Arizona, Tucson, AZ, 85721, USA}
\affiliation{Center for Astrophysics \text{\textbar} Harvard \& Smithsonian, 60 Garden Street, Cambridge, MA 02138, USA}

%%%%%%%%%%%%%%%%%%%%%%%%%%%
%  Abstract                                                                  %
%%%%%%%%%%%%%%%%%%%%%%%%%%%

\begin{abstract}

We simulate the space environment around AU Microscopii b and the interaction between the magnetized stellar wind with a planetary atmospheric outflow for ambient stellar wind conditions and Coronal Mass Ejection (CME) conditions. We also calculate synthetic Ly$\alpha$ absorption due to neutral hydrogen in the ambient and the escaping planetary atmosphere affected by this interaction. We find that the Ly$\alpha$ absorption is highly variable due to the highly-varying stellar wind conditions. A strong Doppler blue-shift component is observed in the Ly$\alpha$ profile, in contradiction to the actual escape velocity observed in the simulations themselves. This result suggest that the strong Doppler blue-shift is likely attributed to the stellar wind, not the escaping neutral atmosphere, either through its advection of neutral planetary gas, or through the creation of a fast neutral flow via charge exchange between the stellar wind ions and the planetary neutrals. Indeed, our CME simulations indicate a strong stripping of magnetospheric material from the planet, including some of the neutral escaping atmosphere. Our simulations show that the pressure around close-in exoplanets is not much lower, and may be even higher, than the pressure at the top of the planetary atmosphere. Thus, the neutral atmosphere is hydrodynamically escaping with a very small velocity ($<15~km~s^{-1}$). Moreover, our simulations show that an MHD treatment is essential in order to properly capture the coupled magnetized stellar wind and the escaping atmosphere, despite of the atmosphere being neutral. This coupling should be considered when interpreting Ly$\alpha$observations in the context of exoplanets atmospheric escape.
    
\end{abstract}


\section{Introduction} \label{sec:intro}

The orbital period vs. planetary radius distribution of confirmed Neptune-size exoplanets reveals a gap at periods less than about 4 days where no ``hot Neptunes'' are found. This ``Neptunian Desert" has been explained by a very high mass loss rate that Neptune-size planets experience at these orbits, leading to the complete evaporation of their gaseous envelope and leaving behind a bare terrestrial-size core. Planets with a much larger size at these orbital distances (hot Saturns and hot Jupiters) seem to have sufficient mass to sustain their gaseous envelope despite their high mass loss rates \citep[e.g.,][]{Szabo11,Beauge13,Batygin16,Lundkvist16,Matsakos16,Mazeh16,Owen18}. The evaporation of the gaseous envelopes in hot Neptunes occurs so quickly that it is very hard to find observational evidence to support the proposed mechanism. The recently discovered exoplanet AU Microscopii b \citep[AU Mic~b hereafter,][]{Plavchan20,Martioli21} offers a unique opportunity to investigate a short-orbit, Neptune-size planet as it undergoes strong evaporation on its way to losing most of its envelope (AU~Mic~b also co-exists with a debris disk). 

Evaporation of gaseous envelopes (the primary, mostly hydrogen atmosphere) of short-orbit planets has been investigated using H~I Ly$\alpha$ observations of several hot Jupiters in absorption during transits. These observations revealed conspicuous blue Doppler shifts in the line, indicating potentially strong outflows of neutral hydrogen from the planet \citep[e.g.,][]{Vidal-Madjar2003,Lecavelier2004,Linsky2010}. The evaporative outflows have been attributed to strong stellar ionizing radiation that leads to hydrodynamic escape \citep[``Parker Wind",][]{Parker:1958} of the neutral hydrogen \citep[e.g.,][]{Erkaev2007,Murray-Clay2009,Owen2012,Shaikhislamov2014}. More recently, similar signatures have been observed in the Helium 10830\AA{} line \citep[e.g.,][]{Spake2018,Oklopcic2018-1,Oklopcic2018-2}. The importance of the latter is that this line can be observed from the ground and it does not get absorbed by the ISM, in contrast to the Ly$\alpha$ line which can only be observed from space due to geocoronal absorption by neutral H in the Earth's atmosphere.

The Ly$\alpha$ and He observations indicate very high mass loss rates of $10^8-10^{10}~g~s^{-1}$. In some cases the blue Doppler shift in Ly$\alpha$ line shows speeds of $100~km~s^{-1}$ or more. Even if the hydrodynamic escape is very strong due to copious EUV irradiation, it still cannot fully explain the high velocity of the escaping neutral hydrogen, and it is even harder to explain in the case of the heavier helium atoms \citep[e.g.,][]{Oklopcic2018-1}.

The high Doppler shift velocity has recently been investigated by \cite{McCann19}. Using a hydrodynamic simulation of the interaction between the stellar wind and the escaping atmosphere, they showed that the stellar wind can sweep up the escaping material and accelerate it to the observed velocities. They also discussed the possibility that the high velocity can be attributed to charge-exchange between the ionized, fast stellar wind ions and the neutral, slow escaping hydrogen atoms, but they did not include this process in their simulation. The interaction between the escaping atmosphere and a generic stellar wind has also recently been studied by \cite{Shaikhislamov2016,Carolan20,Kubyshkina2022} using a hydrodynamic model. They have found that a stronger stellar wind reduces the mass loss rate by a factor of 2. \cite{Hazra2021} used a similar hydrodynamic approach to study the impact on the escaping atmosphere of HD 189733 by stellar CME and flares. They found that the high blue-shifted velocity may be attribute to CMEs.

Both the simulations of \cite{McCann19} and \cite{Carolan20} described the stellar wind - planetary outflow interaction using hydrodynamic models, where the magnetic field and electromagnetic forces have been neglected. However, it is known from the interaction of the solar wind with planetary atmospheres in the solar system that this interaction is controlled by the magnetic field - plasma interaction, even in the case of non-magnetized planets, where a magnetosphere can be induced by the stellar wind pushing against the planet's upper atmosphere \citep{Kivelson.Russell:95,Heliophysics09,SpacePhysics16}. Therefore, the ability of hydrodynamic simulations to properly capture the wind - atmosphere interaction is limited.  \citet{Ekenback2010,Kislyakova2014,Kislyakova2019} used direct Monte-carlo simulations that included the effect of the magnetic field. \citet{Kislyakova2019} used an MHD treatment to set the lower boundary conditions. These detailed simulations, which included photoionization and charge exchange, showed that the observed high velocity Doppler shift could be attributed to Energetic Neutral Atoms (ENA) created via stellar wind - escaping neutrals charge exchange \citep[see also][]{Holmstrom2008}. The focus in these simulations was to fit a particular set of parameters to the observed signal at the time of the observation.

More recently, \citet{Harbach21} performed MHD simulations of the stellar wind interacting with an atmospheric outflow of a TRAPPIST 1e-like planet. They showed how the Ly$\alpha$ line profile can change dramatically on timescales as short as an hour as a function of the stellar wind conditions at different orbital phases. The stellar wind conditions themselves can vary by a few orders of magnitude along the orbits of short-period exoplanets \citep[e.g.,][]{Garaffo16,Garraffo17}. 

In this paper, we extend the work by \cite{Harbach21} to predict how the Ly$\alpha$ observation varies as a function of the stellar wind in AU Mic b. We simulate the stellar wind conditions using an MHD model for the stellar corona and stellar wind, and use the results to drive an MHD model that simulates the stellar wind - magnetosphere interaction. The results of the latter are used to investigate the Ly$\alpha$ line profile as a function of the planetary orbital phase. In addition to the effect of the ambient Stellar Wind (SW), we also investigate the effect of the Coronal Mass Ejection (CME) conditions,  obtained from detailed CME simulations by \cite{AlvaradoGomez22}, on the planetary outflow and its associated Ly$\alpha$ signature. 

We discuss the model setup and the input parameters in Section~\ref{sec:model}, and describe the results in Section~\ref{sec:results}. In Section~\ref{sec:discussion}, we discuss the consequences of our modeling work for future observations of atmospheric escape in AU Mic b, and we conclude our findings in Section~\ref{sec:conclusions}.

\section{Model Description} \label{sec:model}

\subsection{Background Stellar Wind Model} \label{sec:backgroundSWmodel}

We use the Alfv\'en Wave Solar Model \citep[AWSoM,][]{Vanderholst14} to obtain the SW conditions near AU~Mic~b. The model solves the non-ideal magnetohydrodynamic (MHD) equations, which include Alfv\'en waves coronal heating and SW acceleration, as well as thermodynamic terms, on a spherical grid between the base of the stellar corona and $200R_\star$. The assumed stellar parameters in the simulations are: $R_\star=0.75R_\odot$, $M_\star=0.5M_\odot$, and $P_{rot}=4.85~days$ \citep{Kiraga12,Plavchan20}, and the smallest grid size near the inner boundary is $0.025R_\star$ in the radial direction and $1.4~degree$ in the latitudinal/azimuthal directions. 

The inner boundary conditions (which essentially define the final solution) are constrained by surface magnetic field data in the form of a ``magnetogram''. Magnetogram data for AU~Mic has been made available from \cite{Kochukhov20} and from \cite{Klein21}, where the former also proposed the presence of an additional dipole component of about $2~kG$. We investigated test simulations for both \cite{Kochukhov20} and \cite{Klein21} magnetograms but found the results to be similar, especially in terms of the variation of conditions along the orbit of AU~Mic~b \citep[see][for a more complete description of the magnetograms and resulting simulated wind conditions]{AlvaradoGomez22}. Since the main focus of this work is the variation of the absorption profile due to SW variations, we focus here on the SW parameters obtained from the simulations driven by the magnetogram derived by \cite{Klein21}.

Once the inner boundary conditions are specified, the initial condition for the three-dimensional magnetic field is calculated using the potential field approximation \citep{Altschuler69}. The model then solves the non-ideal MHD equations, where it accounts for the Alfv\'en waves coronal heating and wind acceleration, as well as coronal thermodynamics, electron heat conduction and radiative cooling, until a steady-state is obtained. The final steady-state represents the state of the stellar corona and SW during the time that the magnetogram data had been obtained. This approach to obtain the stellar corona and SW solution has been extensively used by e.g., \cite{Cohen2011,Cohen2014,AlvaradoGomez16,doNascimento16,Garaffo16,Garraffo17,Dong17,Vidotto17,Dong18,Kavanagh21}. We refer the reader to those references and to \cite{Vanderholst14} for the complete model description. From the three-dimensional steady-state coronal solution, the parameters along the orbit of the planet are extracted and are used to drive the global magnetosphere model (see Section~\ref{sec:GM}). The parameters are extracted with a resolution of 3 degrees along the circular orbit (about 3.5 hours). 

\subsection{CME Model} \label{sec:CMEmodel}

In addition to the steady-state SW conditions, we calculate Ly$\alpha$ absorption profile variations due to a CME that hits the planet AU~Mic~b. The CME is obtained in the coronal model by superimposing an unstable flux-rope onto the steady state solution. Due to its instability, the flux-rope erupts and its evolution, propagation, and interaction with the ambient SW is simulated in a time dependent manner. The CME conditions near the planet are recorded as a function of time, and they are also used to drive the global magnetosphere model (similar to the orbital SW conditions). Here, we simulate 90 minutes of the CME as it passes through the location of AU Mic b.

The prescription for the unstable flux-rope is given by \cite{Titov99} and it has been used for simulations of CMEs at Earth \citep[e.g.,][]{Jin13} and for simulations of stellar CMEs \citep{AlvaradoGomez19}. The bipolar flux-rope parameters include its location on the photosphere, tilt angle, separation length, internal radius, total plasma mass, and stored energy. The latter is controlled by a current term in the flux-rope formalism. The location of the flux-rope was determined using data from \cite{Wisniewski19}, while the CME mass, $M_{CME}=10^{20}~g$, and energy, $E_{CME}=10^{36}~ergs$, were fit to the best current CME candidate event on AU~Mic \citep[][see also \citealt{Katsova99,Cully1994}]{Moschou_2019}. The complete description of the CME simulation can be found in \cite{AlvaradoGomez22}.

\subsection{Global Magnetosphere Model} \label{sec:GM}

In order to study the impact of the stellar environment conditions on the Ly$\alpha$ profile, we use the {\it BATSRUS} Global Magnetosphere (GM) MHD code \citep{Powell:99,Toth2005,Toth.etal:12}. The model is constructed with a spherical inner boundary at the planetary surface, and a cartesian grid that extends between $+100R_p$ to $-200R_p$ along the star-planet line ($X_{GSE}$, day to night side), and between $+130R_p$ to $-130R_p$ in the other directions ($Y_{GSE}$ and $Z_{GSE}$). Here, GSE is the Geocentric solar ecliptic coordinate system \citep{Kivelson.Russell:95}. We adopt a planetary radius $R_p=25,930$~km \citep[$0.35R_J$, $4R_E$,][]{Martioli21}. The grid size near the inner boundary is $\Delta x=0.1R_p$.

The space environment conditions for the orbital SW or the time-dependent CME propagation near the planet are imposed on the outer face of the GM domain which faces the star. The conditions are given as time-dependent upstream conditions, and the time-dependent GM simulation provides the response of the escaping atmosphere and magnetosphere to variations in these upstream conditions. 

We are interested in capturing the variations of the escaping atmosphere and its observable signature (i.e., the Ly$\alpha$ profile) due to the changing upstream conditions. Thus, we impose a prescribed inner boundary condition for the plasma density and temperature which results in a strong outflow due to the pressure gradient. We stress that we do not attempt to simulate an actual photoevaporative outflow and its acceleration at lower parts of the atmosphere. We do not even specify an outflow velocity (set to zero) at the boundary, and only specify temperature and density. Our approach enables us to obtain an outflow that is strong enough to interact with the incoming SW/CME, as well as to produce a Ly$\alpha$ absorption signature. The prescribed boundary conditions are number density of $n=3\cdot 10^8~cm^{-3}$ and temperature of $T=10^4~K$, resulting in an outflow mass-loss rates shown in Table~\ref{table1}. The mass-loss rates are calculated over a sphere just slightly above the spherical inner boundary (located at $1.2R_p$), and are of the order of $10^{10}~g~s^{-1}$. This falls within the low range of mass-loss rates of $10^{10}-10^{12}~g~s^{-1}$ predicted in hot Jupiters \citep{Owen19}. Of course, the mass loss rate is controlled mostly by the base density value used in the simulations. 

\begin{figure*}[h!]
\centering
\includegraphics[width=6.75in]{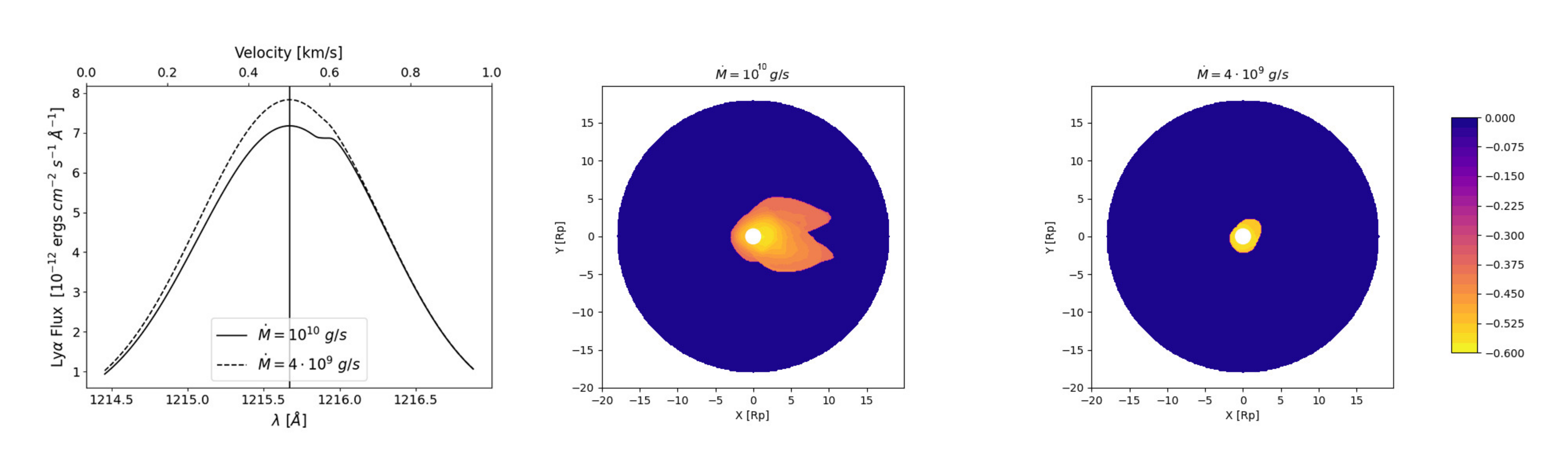}
\caption{Ly$\alpha$ line absorption (left) and absorption images (middle and right) for planetary mas-loss rate of $4\cdot 10^9$ and $10^{10}~g~s^{-1}$.}
\label{fig:MassLossComparison}
\end{figure*}

We stress that our goal here is to drive a strong outflow to investigate the impact of the stellar wind on the outflow and its Ly$\alpha$ signature. We do not aim to predict an actual observed profile for AU Mic b. Thus, we apply a prescribed pressure at the inner boundary which leads to an outflow from it. This is different from one-dimensional models of hydrodynamic escape that use the energy-limited stellar radiation to heat the base of the atmosphere and to create the high base pressure along with a very low pressure at the top boundary. \cite{Kubyshkina2018} performed a grid of hydrodynamic models for different exoplanets. AU Mic b is probably most similar to the $Pd2$ case from \cite{Kubyshkina2018}, except that here we set the base temperature to $10^4~K$, with a mass-loss rate of $10^{10}~g~s^{-1}$,  while Case $Pd2$ has a temperature of $4370~K$ and a mass-loss rate of $2\cdot 10^{9}~g~s^{-1}$. As a sanity check, we re-run our Case 1 using the temperature of $4370~K$ and we obtained a mass-loss rate of $4\cdot 10^{9}~g~s^{-1}$, which is more or less consistent with \cite{Kubyshkina2018}. Since we aim to demonstrate that the Ly$\alpha$ signature is sensitive to stellar wind variations, overestimating the outflow here does not matter, since if the mass-loss rate would have been lower, it would be even more sensitive to the stellar wind variations. Figure~\ref{fig:MassLossComparison} compares the Ly$\alpha$ absorption images and line profiles for the two prescribed mass-loss rates. It can be seen that a lower mass-loss rate results in a less visible absorption (see the following section for a detailed description of how these images are produced).

A planetary magnetic field of $B=-0.3~G$ (Earth-like) is assumed in all cases. We leave the study of the impact of different planetary magnetic field strengths on the planetary outflow for a future investigation (see Section~\ref{sec:MagneticFieldRole}).

\subsection{Synthetic Ly$\alpha$ Profile} \label{sec:syntheticLy}

We follow the formalism by \cite{Tasitsiomi06}, \cite{Khodachenko17}, and \cite{Harbach21} in order to obtain the Ly$\alpha$ synthetic observables. First, we define a line-of-sight (LOS) which points in the positive $X_{GSE}$ direction, which extends from the back of the GM domain (behind the planet) towards the star. We calculate the intensity of the Ly$\alpha$ emission, $I_{Ly}(v)$, which comes out from the star and passes through the planetary atmosphere, for a given velocity, $v$:
\begin{equation}
\label{EqLy}
I_{Ly}(v)=\int \int I_{Ly0} (y,z,v) \exp{[-\tau(y,z,v)]}dydz,
\end{equation}
with $I_{Ly0} (y,z,v)$ being the background stellar intensity, and $\tau(y,z,v)$ being the optical depth. A strong absorption leads to an increase in the optical depth, and a reduction in the Ly$\alpha$ intensity at the observation point. The optical depth is obtained by integrating the column density and the Ly$\alpha$ cross section, $\sigma_{Ly(v)}$:
\begin{equation}
\tau(y,z,v)=\int_{LOS} 0.5\cdot N_H(x)\sigma_{Ly}(v)dx.
\end{equation}
Here $N_H(x)$ is the local density at the cell along the LOS, and we assume a 50\% ionization of the gas, hence the factor of 0.5 \citep{Owen2016}. 

The cross section is approximated as:
\begin{eqnarray}
\sigma_{Ly}(v) = 5.8\times10^{-14} \sqrt{10^4K/T(K)}\cdot\exp{(-b^2)}\\
+2.6\times10^{-19}\left[ \frac{100km~s^{-1}}{v-v_x} \right]^2 q(b^2),
\end{eqnarray}
where
\begin{equation}
b=\frac{v-v_x}{\sqrt{2kT/m_p}}
\end{equation}
with $k$ is the Boltzmann constant, and $m_p$ is the proton mass. The coefficient $q$ is given by: 
\begin{equation}
  q(b^2)=
  \begin{cases}
  	\frac{21+b^2}{1+b^1}z(b^2) (0.1117+z(b^2)) & z(b^2)>0 \\
  	\times[4.421+z(b^2)(5.674z(b^2)-9.207)]& \\
	0 & z(b^2)<0,
  \end{cases}
\end{equation}
with
\begin{equation}
z(b^2)=\frac{b^2-0.855}{b^2+3.42}.
\end{equation}

The stellar background Ly$\alpha$ flux, $I_{Ly0}$, is assumed to be a Gaussian with a full width at half maximum (FWHM) of $148.6~km~s^{-1}$, centered at $1215.67\AA$, and peaking at $10^{-11}~ergs~cm^{-2}~s^{-1}$ \citep{Schneider2019}. 

For each of the GM MHD solutions, we perform the LOS integration for each line on the grid in the $X_{GSE}$ direction for a given velocity. The velocity grid runs from $-300~km~s^{-1}$ to $+300~km~s^{-1}$ with a resolution of $ 10~km~s^{-1}$. Each LOS integration provides a pixel in a $Y_{GSE}$-$Z_{GSE}$ image. At the end of the procedure, we obtain an intensity image for every velocity, $I_{Ly}(y,z,v)=exp{(-\tau)}$. The image for each velocity is multiplied by a background stellar disk that covers a circle of $18R_p$ (AU Mic's radius in units of AU Mic b's radius), and the result is integrated over the image to give the total flux at a given velocity (or Doppler shift), i.e., Eq.~\ref{EqLy}. Multiplying this flux with the background stellar flux Gaussian gives the absorption at each velocity, and the Ly$\alpha$ line Doppler profile.

\section{Results} \label{sec:results}
\subsection{Background Stellar Wind Conditions} \label{sec:backgroundSW}

\begin{figure*}[h!]
\centering
\includegraphics[width=7.in]{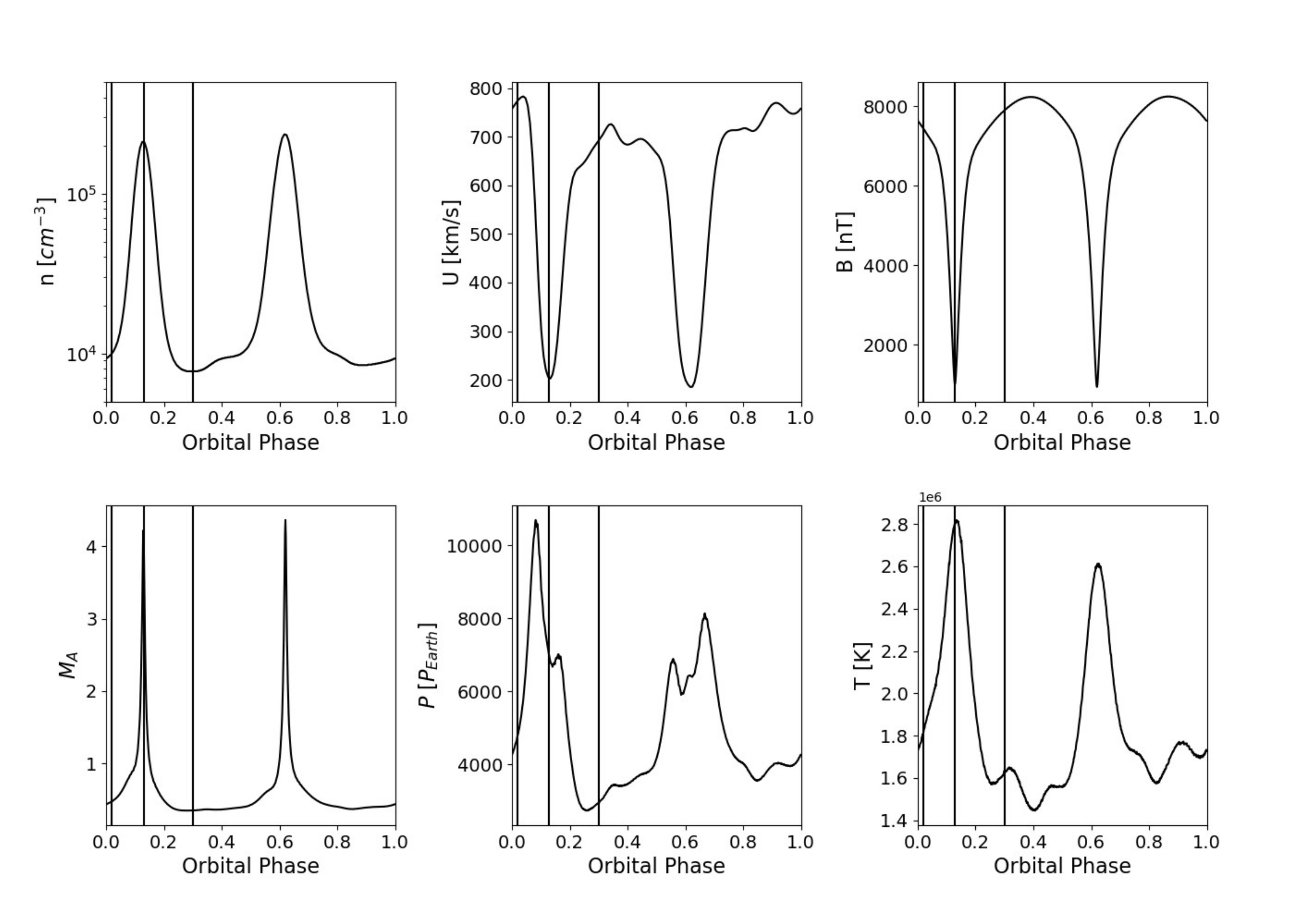}
\caption{SW parameters as extracted from the {\it AWSOM} solution along the orbit of AU Mic b and used to drive the GM model. Plots are for the SW number density (top-left), SW speed (top-middle), SW magnetic field strength (top-right), and SW Alfv\'enic Mach number (bottom-left), SW dynamic pressure (bottom-middle), and SW temperature (bottom-right). SW conditions for cases 1-3 are marked by the vertical black lines (left to right, respectively).}
\label{fig:Orbitconditions}
\end{figure*}

Figure~\ref{fig:Orbitconditions} shows the SW conditions along the orbit of AU~Mic~b. It shows the orbital variations of the number density, $n$, the SW speed, $u$, the SW magnetic field strength, $B$, the SW Alfv\'enic Mach number, $M_A=u/u_A$, the SW dynamic pressure, $p$ (normalized to typical solar wind conditions at 1~AU with $n=5~cm^{-3}$ and $u=500~km~s^{-1}$), and plasma temperature, $T$. The local Alfv\'en speed is  $u_A=B/\sqrt{4\pi\rho}$, with $\rho=nm_p$ being the mass density, and $m_p$ the proton mass. The SW velocity and magnetic field vectors, as well as the number density and temperature are used to drive the GM model. The stellar wind conditions vary strongly with azimuthal angle. In particular,  AU~Mic~b will reside in a sub-Alfv\'enic regime for most of the orbit, but transitions into a super-Alfv\'enic regime twice during the crossing of the astrospheric current sheet. 

The SW speeds are not so different from those observed near the Earth. However, the number density and the magnetic field strength are orders of magnitude higher due to the close proximity of the orbit to the star (typical values for the Earth are $n=1-100~cm^{-3}$ and $B=1-500~nT$, for ambient solar wind and CME conditions\footnote{https://cdaweb.gsfc.nasa.gov/istp\_public/}). The orbital period of AU~Mic~b is about $8.5~d$ \citep{Plavchan20,Martioli21}. Thus, the crossing of the astrospheric current sheet (low to high density and high to low magnetic field/wind speed) takes about two days, and the transition between the sub- to super-Alfv\'enic SW takes less than a day. \cite{Cohen2014} have shown that such quick transitions may lead to a rapid global change in the magnetospheric structure from an Alfv\'en wings \citep{Neubauer80,Neubauer98}, Io-like topology to a stretched, Earth-like topology, which leads to strong variations in the SW-planetary outflow interaction. 

Here, we focus on three azimuthal angles in the simulations that we use as representative points along the orbit of AU~Mic~b. We investigate a case of the sub-Alfv\'enic region at the beginning of the orbit (Case 1), a case of the super-Alfv\'enic point (Case 2), and a case of the second sub-Alfv\'enic region (Case 3). We assume that all other orbital locations behave similarly to these locations.

\begin{figure*}[h!]
\centering
\includegraphics[width=6.75in]{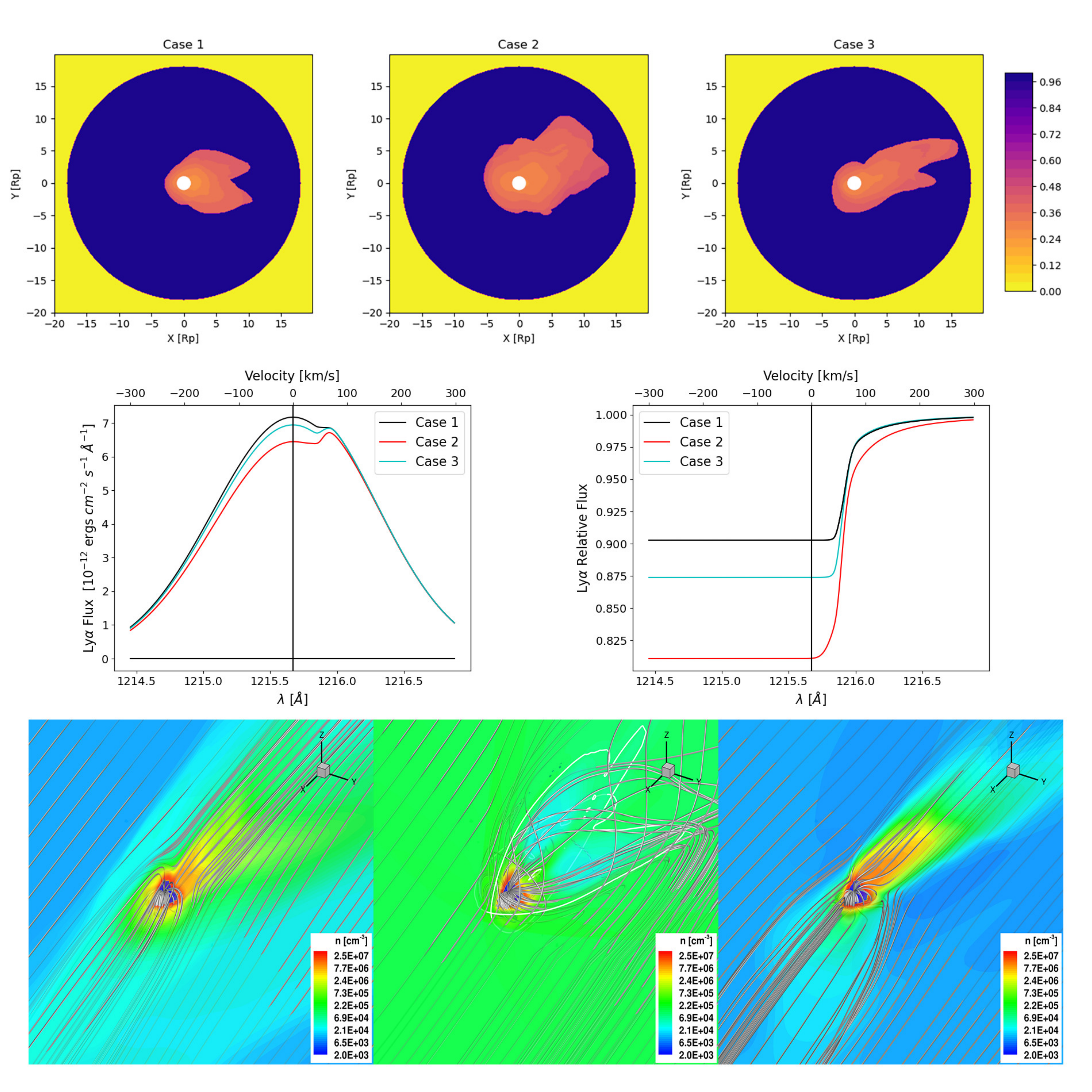}
\caption{Top: Absorption ($e^{-\tau}$) images for the three cases integrated over all velocities. Middle: Ly$\alpha$ Doppler-shift profile for the three cases (left), and the ratio of these profiles to the un-absorbed profile (right).) Bottom: the three-dimensional solution near the planet for Cases 1-3 (left, middle, right, respectively). Color contours are for the number density where selected magnetic field lines are also shown. The solid white line marks the Alfv\'en surface (seen only in the super-Alfv\'enic Case 2).}
\label{fig:CasesResults}
\end{figure*}

Figure~\ref{fig:CasesResults} shows the results for Cases 1--3. The top panel shows the mid-transit absorption ($e^{-\tau}$) images integrated over all velocities. It can be seen that the Ly$\alpha$ absorption corresponds to the cooler (under 50,000K), denser material that occupies the inner regions of the planetary magnetosphere. The middle panel of Figure~\ref{fig:CasesResults} shows the Ly$\alpha$ Doppler profile for the three cases (left), and the ratio of their absorbed to the non-absorbed (stellar background) profiles (right). These plots show significant absorption in the blue wing of the profile, ranging between 10-20\%, and peaking at velocities between $-40$ to $-100~km~s^{-1}$ ($1215.25-1215.5$\AA). 

In the bottom panel of Figure~\ref{fig:CasesResults}, we show the corresponding three-dimensional density and magnetic field structure of the planetary magnetosphere for the three orbital epochs. It shows that initially (Case 1), the planet resides in a sub-Alfv\'enic SW with two magnetospheric lobes extended towards and away from the star, with an angle that depends on the particular SW magnetic field vector at that time. A relatively high density is seen close to the planet due to the planetary outflow from the inner boundary. When the planet moves to the super-Alfv\'enic SW (Case 2, current sheet crossing), the SW density increases, stretching the magnetospheric lobes behind the planet, forming a bow shock at the day side of the magnetosphere, and pushing against the planetary outflow. When the planet moves back to the lower density, sub-Alfv\'enic SW region (Case 3), the initial structure is recovered (even though Cases 1 and 3 are not identical).

\begin{figure*}[h!]
\centering
\includegraphics[width=5.5in]{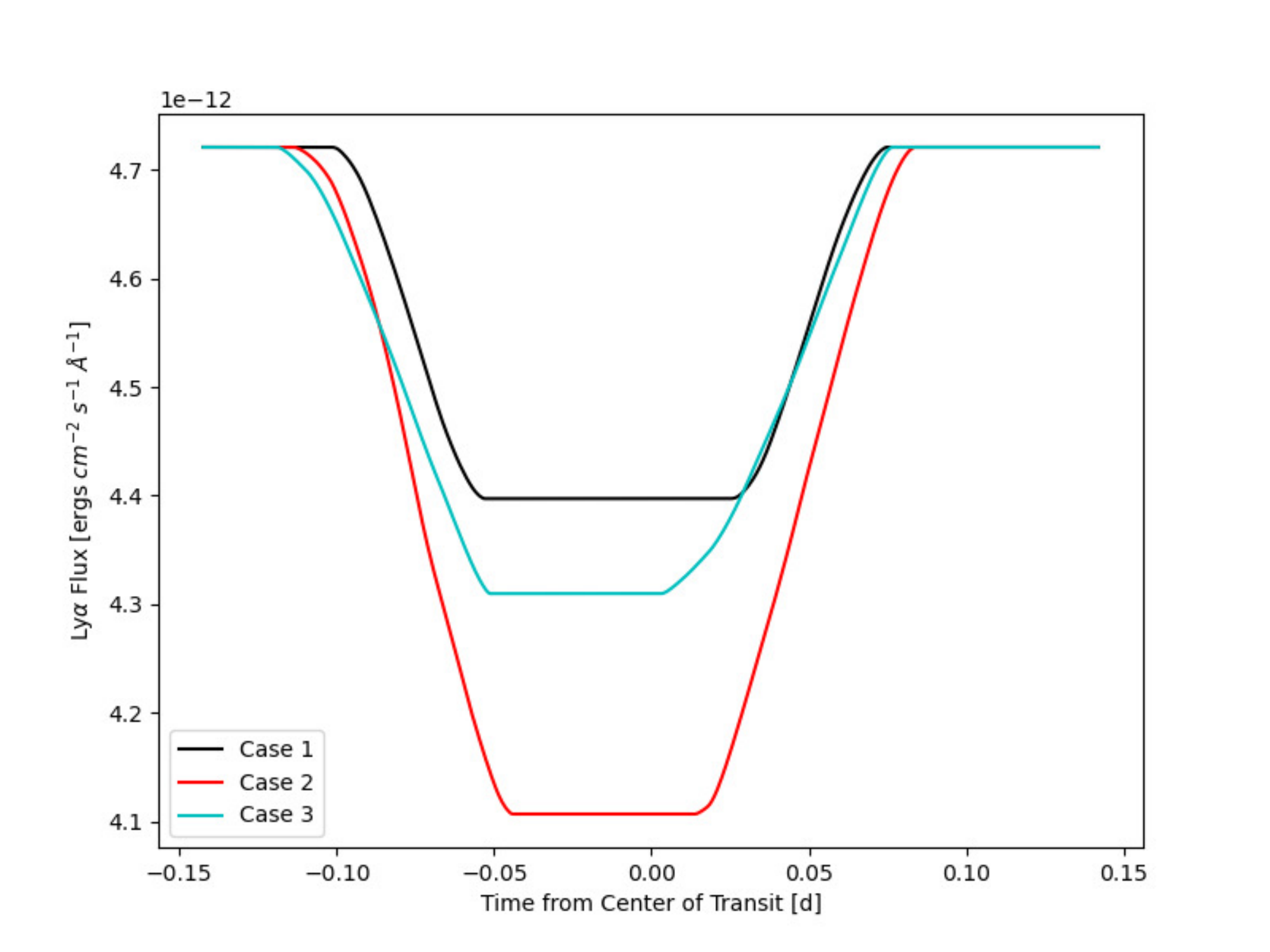}
\caption{Synthetic transit profile of the Ly$\alpha$ flux for Cases 1-3.}
\label{fig:Lightcurve}
\end{figure*}

Figure~\ref{fig:Lightcurve} shows the predicted Ly$\alpha$ lightcurve. It shows the Ly$\alpha$ flux integrated over all velocities as the planet passes in front of the stellar disk. Our predicted transit lasts for about 4-5h, which is consistent with e.g., \cite{Plavchan20}. Figure~\ref{fig:Lightcurve} and the middle panel of Figure~\ref{fig:CasesResults} clearly shows that the Ly$\alpha$ transit changes significantly along the orbit due to the different SW conditions that the planet experiences. The blue wing and the overall absorption is less than 10\% in Case 1, then it increases to almost 20\% in Case 2, then it decreases again to about 13-14\% in Case 3. These absorption variations occur on a time scale of less than a day, and they are clearly due to the SW variations as our planetary outflow driver at the inner boundary is uniform and steady. It is also clear from the top panel of Figure~\ref{fig:CasesResults} that the SW conditions change the overall (density) structure of the inner magnetosphere, which is the region that is also responsible for the absorption of the stellar Ly$\alpha$ radiation. 

\subsection{CME Conditions} \label{sec:CME}

\begin{figure*}[h!]
\centering
\includegraphics[width=7in]{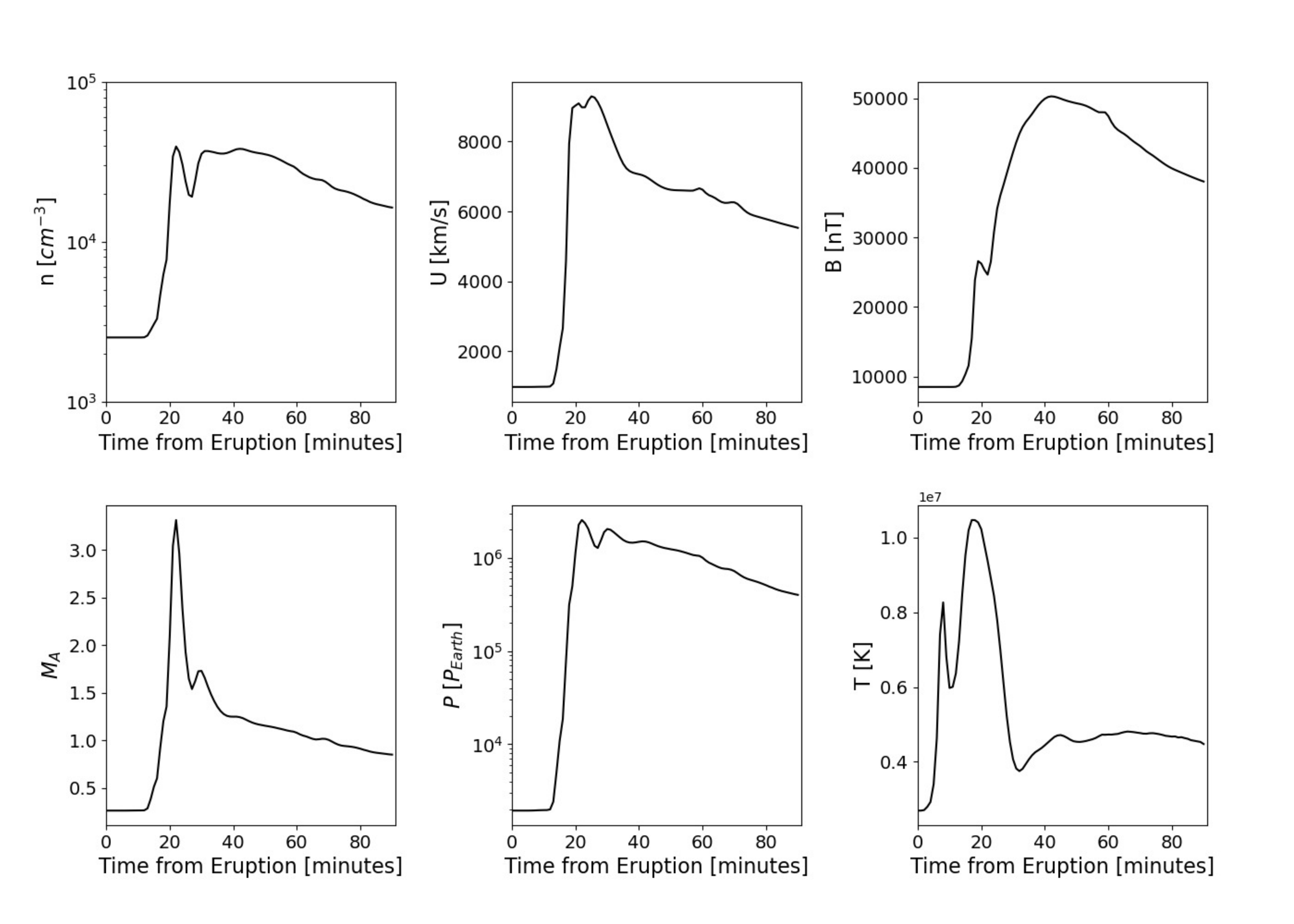}
\caption{SW parameters as extracted from the {\it AWSOM} solution near the planet during the CME event (display is similar to Figure~\ref{fig:Orbitconditions}). In this case, the exoplanet (and the GM domain) are assumed to be located at an orbital distance from the star along the star-planet line, facing the center of the CME (similar to an L1 point satellite near the Earth measuring an incoming CME). The CME arrival time to the edge of the GM solution occurs around $t=20~min$. }
\label{fig:CMEconditions}
\end{figure*}

Figure~\ref{fig:CMEconditions} shows the SW conditions for the CME event hitting AU~Mic~b. These conditions were extracted with a 1 minute cadence over the course of 90 minutes. The CME front is clearly seen in the Alfv\'enic Mach number plot, where it arrives at the planet after only about 20 minutes. This is not surprising due to the very fast CME speed of over $8000~km~s^{-1}$. The CME front is also indicated by the sudden increase in density, magnetic field, and speed. The post-front CME conditions occupy the vicinity of the planet for the rest of the simulation.

\begin{figure*}[h!]
\centering
\includegraphics[width=6.75in]{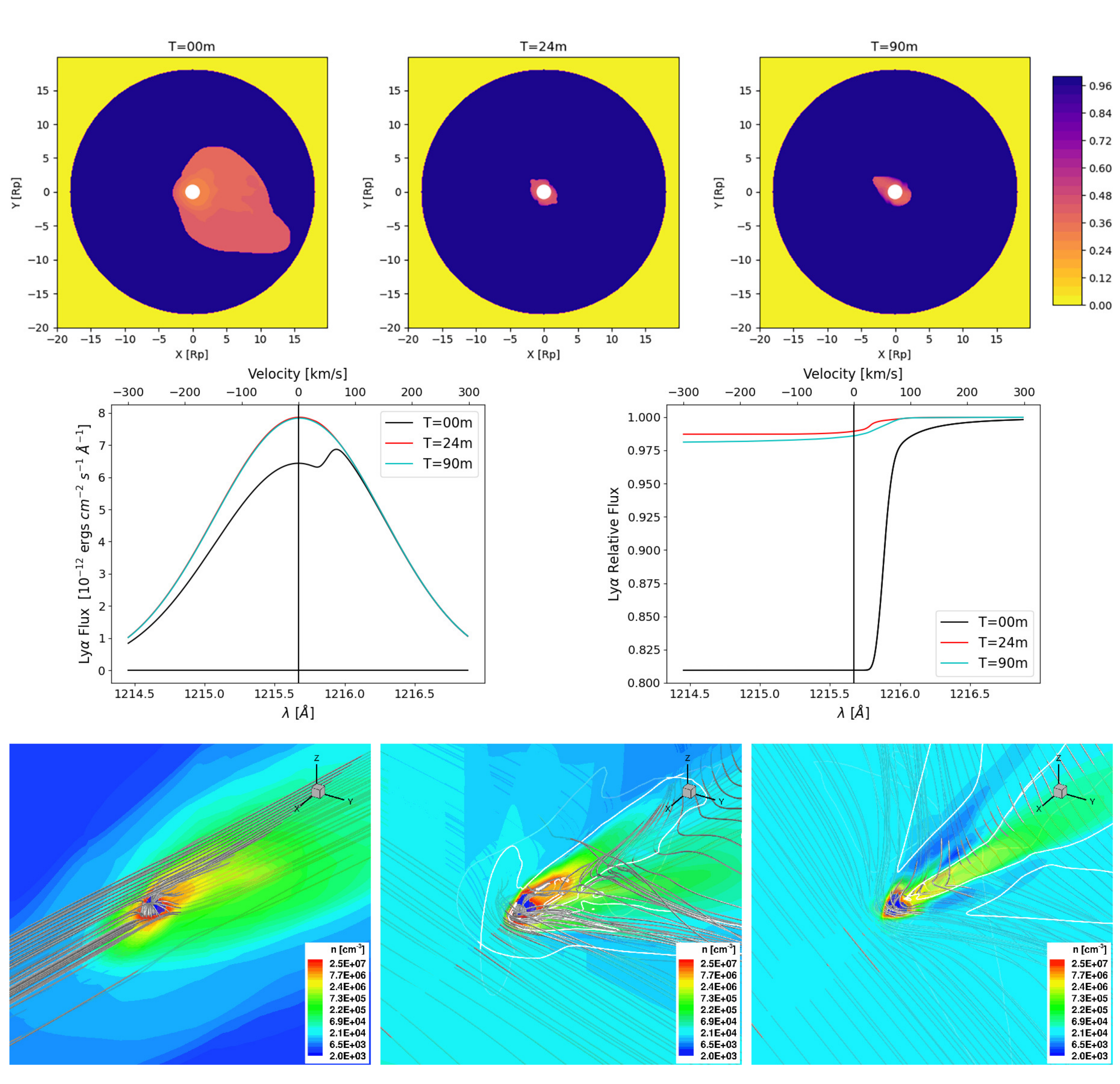}
\caption{Similar display as in Figure~\ref{fig:CasesResults} but for the times of the CME event, showing the CME has pressure-stripped the escaping envelope, with the Ly$\alpha$ absorption greatly reduced in the $T=24$m and $T=90$m panels.}
\label{fig:CMEResults}
\end{figure*}

Figure~\ref{fig:CMEResults} is similar to Figure~\ref{fig:CasesResults} but it shows the results for the CME event at $t=0$, $24$, and $90$~minutes. The choice of $t=24$~min is due to the fact that the CME front arrives at the edge of the GM simulation domain after 20 minutes, but it takes 4 more minutes to see the impact of the CME on the absorption profile when the CME reaches close proximity to the planet. It can be seen that prior to the arrival of the CME at the initial state, a strong absorption occurs in a similar manner to Case 1, as the SW conditions at that time are sub-Alfv\'enic. When the CME arrives after 24 minutes, and throughout the CME event up to $t=90~min$, the signature completely disappears from both the absorption images and the Ly$\alpha$ Doppler profile. The three-dimensional plots at the bottom panel of Figure~\ref{fig:CMEResults} clearly show that the CME plasma takes over most of the simulation domain, replacing the magnetospheric material that occupied the domain prior to the CME arrival. 

\section{Discussion} \label{sec:discussion}

The results presented above highlight the conclusions of \citet{Harbach21}: for close-in planets the stellar wind is instrumental in shaping atmospheric outflows and can dictate the form and variability of absorption signatures. Such variability can be large. Different stellar wind and magnetic field conditions with azimuthal angle imply that strong variations in outflow signatures will occur over an orbit, potentially leading to observable variations in absorption signatures during a single transit on timescales as short as a few hours. 

Regardless of any secular evolution of the stellar surface magnetic field that could lead to changes in the wind conditions and atmospheric outflow absorption signatures, the occurrence of transit-to-transit variations can be expected.

\subsection{Stellar-wind Driven Variations of the Ly$\alpha$ Observables} \label{sec:LyAlphaVariations}

One of the aims of this study is to shed light on the relation between the three-dimensional physical system and the one-dimensional, Ly$\alpha$ signal that is often the only observed diagnostic. In interpreting observed Ly$\alpha$ profiles as an indicator of a strong atmospheric outflow, many assumptions are made regarding the space environment and stellar wind near the planet, the particular modeling approach and assumptions, and the radiative transfer processes. It is important to stress that these interpretations are generally model dependent (or ``assumptions dependent'' on the practical level). Our simulations provide a number of important features that could constrain these assumptions, in the context of the three-dimensional system.

As first proposed by \cite{Vidal-Madjar2003}, an excess of blue-shifted Ly$\alpha$ absorption during transit as compared with the out of transit flux indicates a flow of neutral hydrogen in the direction of the observer. Assuming that the neutral hydrogen is concentrated in the planetary atmosphere, the observed blue-shift of about $100~km~s^{-1}$ indicates a strong outflow of neutral hydrogen from the planet with that speed. The strong absorption also requires a quite high density, with values that are likely to be possible only at the top of the planetary atmosphere (below an altitude of a few thousand km). In contrast, models for neutral hydrogen hydrodynamic escape from hot Jupiters \citep[e.g.,][]{Murray-Clay2009,Owen2012,Tripathi2015,McCann19} predicted escape velocities in the range of $5-20~km~s^{-1}$. 

Despite the high mass-loss rate we obtain in our simulations, the outflow speed near the planet is much less than $20~km~s^{-1}$, which is consistent with the hydrodynamic escape models referenced above. This is far lower than the negative $100~km~s^{-1}$ speed associated with the Ly$\alpha$ observations. We draw a particular attention to an important aspect of the nature of the hydrodynamic escape process. In its essence, hydrodynamic escape is driven by a pressure gradient between the planet and space, and the greater this pressure gradient is the greater the outflow.
%In his original paper on the solar wind, magnetic field and interstellar medium, \cite{Parker:1958} showed the non-zero pressure estimated for the interstellar medium, which contradicted the pressure predicted by a hydrostatic solar atmosphere ($p(r\rightarrow \infty)=0$). 
Indeed, in his original paper on the solar wind, magnetic field and interstellar medium, \cite{Parker:1958} showed that the low pressure estimated for the interstellar medium was consistent with hydrodynamic outflow of a $3\times 10^6$~K solar corona.
Such a setting is assumed in many models for hydrodynamic escape (mentioned in Section~\ref{sec:intro}). In our simulations here, we impose a high pressure at the inner boundary via high density and temperature that were obtained from the literature. If our simulated exoplanet would be located far from the star, then a high escape rate would be self-consistently generated in the simulation via the pressure gradient. However, in the case of close-in exoplanets, space is not that empty (i.e., the pressure is not necessarily low). The ambient SW density near AU Mic b is 3-4 orders of magnitude higher than that near the Earth, and the temperature is approaching the coronal temperature of more than 1MK. Thus, the pressure at ``infinity'' is not much smaller, and in some situations may actually be higher than that of the planet inner boundary. 

\begin{figure*}[h!]
\centering
\includegraphics[width=6.75in]{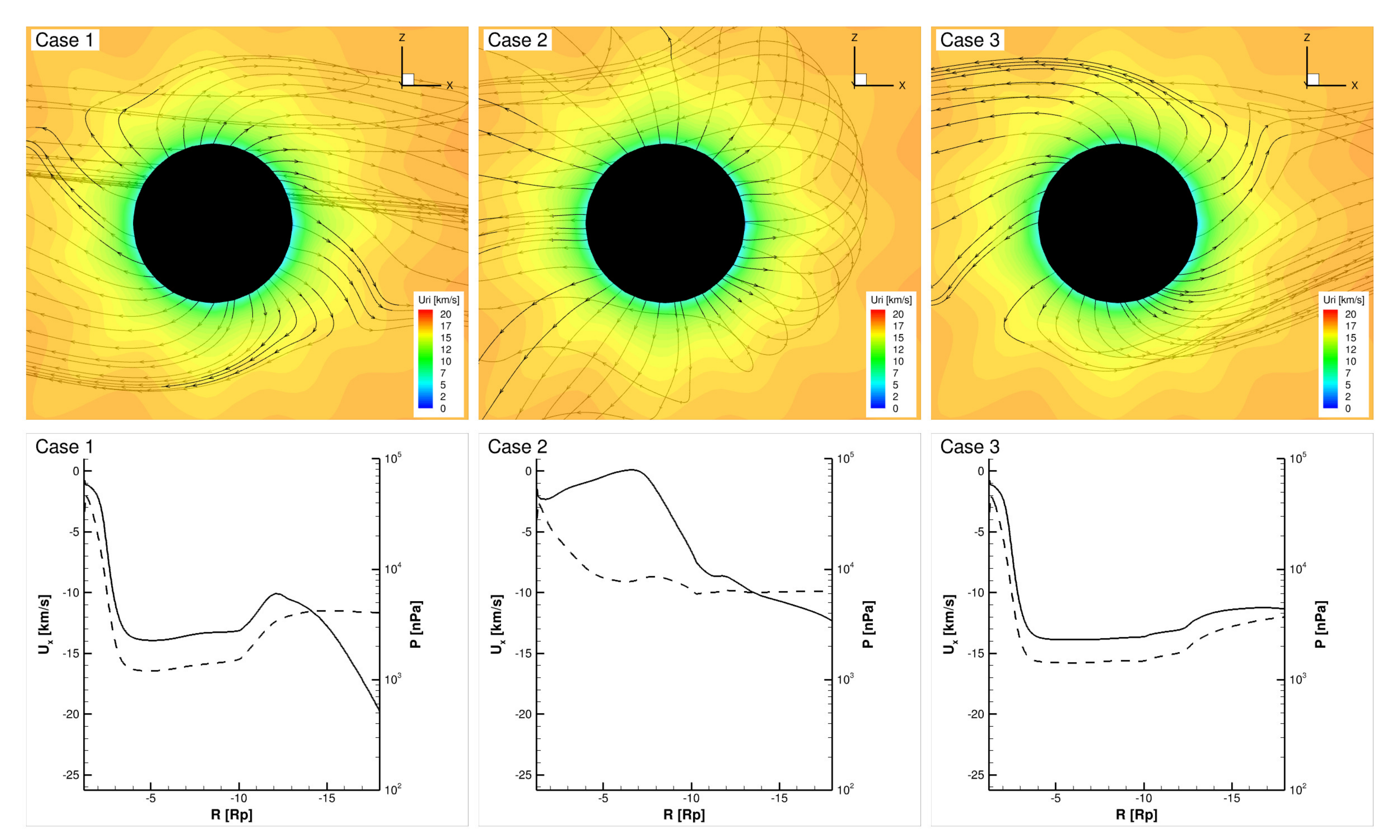}
\caption{Top: color contours of the radial velocity displayed on the $x-z$ plane for Cases 1-3. Velocity three-dimensional streamlines are represented by the black arrow lines. Bottom: Line plots show the LOS ($u_x$) velocity component (solid) and the thermal pressure (dashed) profiles along the negative $\hat{x}$ (night side) line starting from the planet up to $20~R_p$.}
\label{fig:UxP}
\end{figure*}

Figure~\ref{fig:UxP} shows the radial velocity close to the planet (magnitude and streamlines displayed on a $x-z$ slice), together with the $U_x$ velocity component (the velocity component towards the observer) and the thermal pressure profile as a function of distance from the planet. The latter were  extracted from the simulation at the back of the planet (night side) in the negative $\hat{x}$ direction. It can be seen that the velocity close to the planet is much less than even $15~km~s^{-1}$, and it gets to higher negative values far from the planet, indicating a SW plasma, not planetary material. It also shows that the thermal pressure near the planet does not drop much: by an order of magnitude or so in the sub-Alfv\'enic cases (1 and 3) within 3-4 $R_p$, and by an even smaller factor in the super-Alfv\'enic case (2). The weak pressure gradient may be even weaker if we consider the total pressure, which is the sum of the thermal, dynamic, and magnetic pressures.

\begin{figure*}[h!]
\centering
\includegraphics[angle=-90,width=5in]{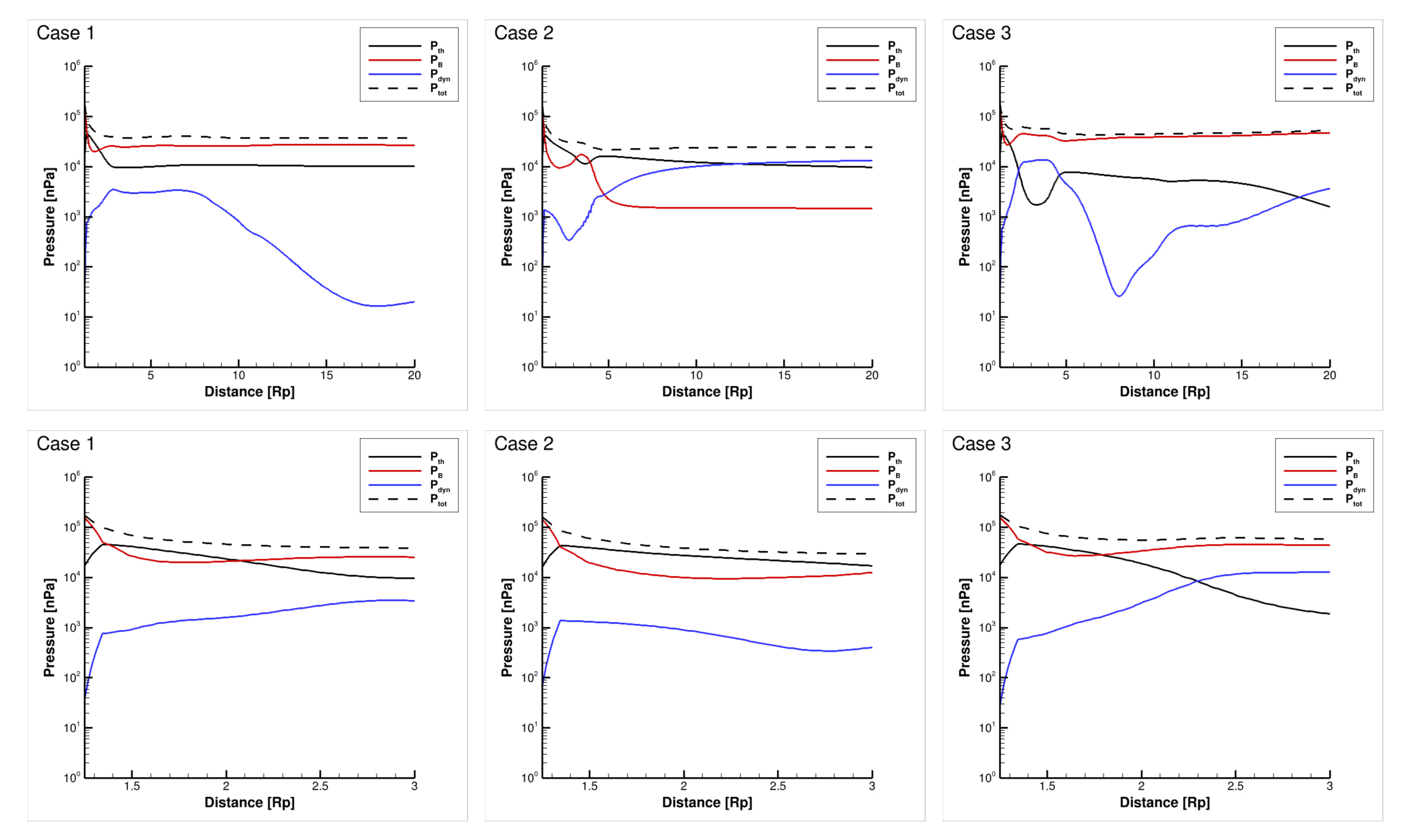}
\caption{Profiles of the different pressure components as a function of radial distance extracted along the sub-stellar line (positive $\hat{x}$, day side) from the planet up to $20~R_p$ (right column), and zoomed up to $3~R_p$ (left column). Plots are for Cases 1-3. Solid lines represent the different pressure components while the dashed line represents the total pressure.}
\label{fig:Cases_pressure}
\end{figure*}

\begin{figure*}[h!]
\centering
\includegraphics[angle=-90,width=5in]{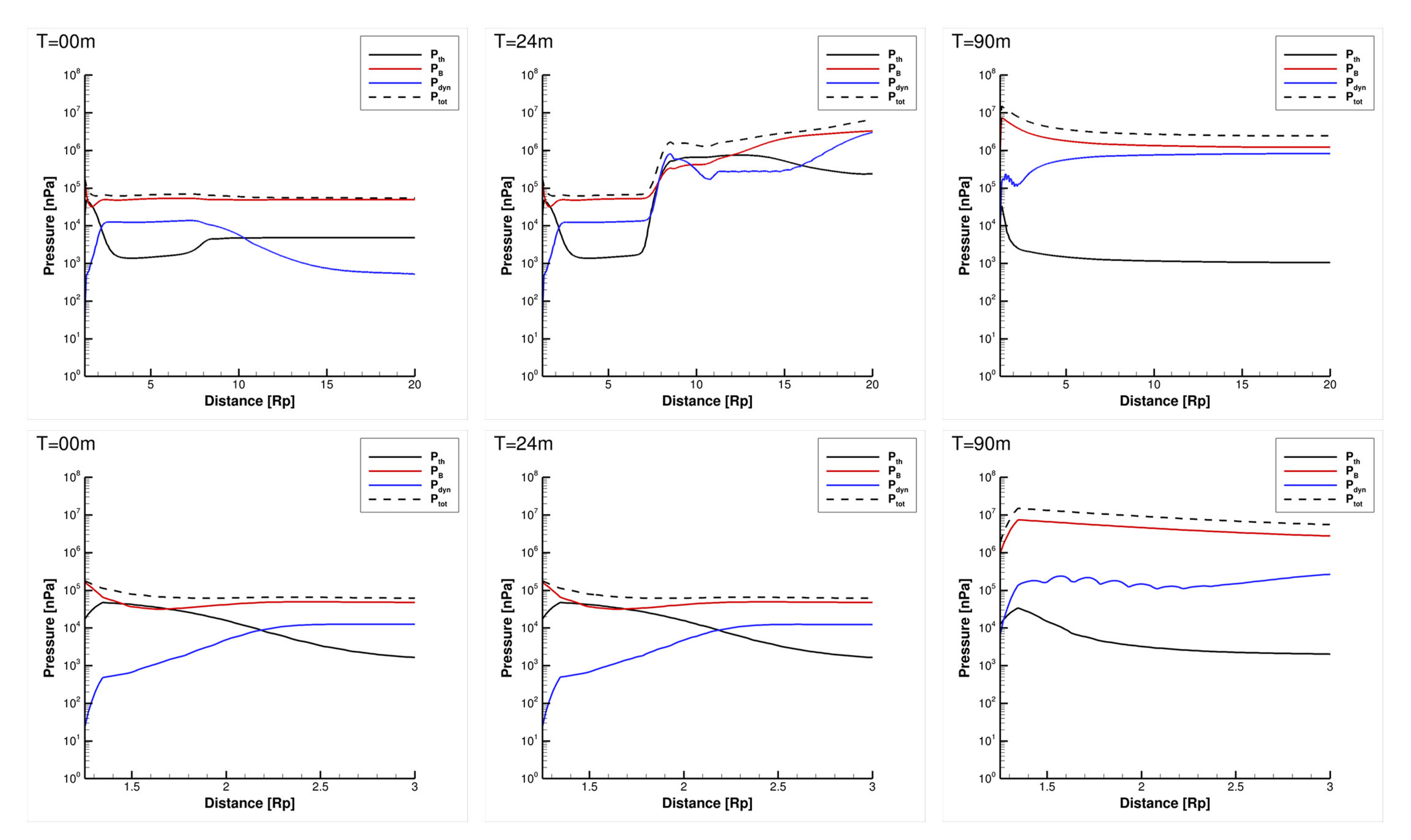}
\caption{Similar to Figure~\ref{fig:Cases_pressure} but showing the profile of the various pressures vs.~ radial distance from the planet for the different times during the CME event close to the planet (left column) and out to $20~R_p$ (right column). The pressure increase due to the CME can be seen at $T=24$m at about $8R_p$.}
\label{fig:CME_pressure}
\end{figure*}

Figures~\ref{fig:Cases_pressure} and \ref{fig:CME_pressure} show the different pressures extracted along the planet-star line (sub-stellar line) for Cases 1-3 and the CME epochs, respectively. The magnetic pressure, which is very small in the SW near the Earth, is much more significant in the case of AU Mic (and close-in planets in general). It is actually the dominant pressure term in some parts. This emphasises the importance of including magnetic effects in studies of  planetary atmospheric escape. When considering the total pressure, the pressure gradient becomes even more moderate than that of the thermal pressure alone.

The reduction of the pressure far from the planet is taken to the extreme in our CME simulations (Figures~\ref{fig:CMEconditions} and \ref{fig:CME_pressure}). Here, it is clearly seen how the increased SW pressure completely shuts down and suppresses the escape of planetary material via reduction of the pressure gradient.
%, and via stripping of the planetary material by the CME. On the one hand, the stripping effect is similar to the escape suppressing by a strong SW, as proposed by \cite{Vidotto20}. On the other hand, 
On the one hand, this is similar to the escape suppressing by a strong SW, as proposed by \cite{Shaikhislamov2016,Vidotto20}. However, the escape suppression clearly occurs due to the overall reduction in the pressure gradient surrounding the planet. In general, our simulations show that for close-in exoplanet simulations of hydrodynamic escape, the upper boundary conditions for the pressure (or the pressure at "infinity") need to account for the rather high SW pressure, including the magnetic pressure.  

\subsection{Mass-loss Due to the CME Event} \label{sec:CMEMassLoss}

One important aspect of CME events on AU~Mic is their possible impact on atmospheric escape. Our GM simulation cannot capture any actual impact on the escape in terms of atmospheric heating or an acceleration of escaping particles \citep[see e.g.,][]{Cohen2014,GarciaSage2017}. However, we can still look at the effect on the outflow of the CME as it passes the planet. 

Our planetary outflow is driven by setting a constant, high pressure at the inner boundary, which is higher than that of the surrounding space, creating a mass-loss rate of about $10^{10}~g~s^{-1}$. However, as the CME arrives, the total pressure in the vicinity of the planet becomes larger than the pressure at the inner boundary, so the outflow is shut down. 

Table~\ref{table1} shows the mass-loss rates during the CME event. Initially, the outflow is $2\times 10^{10}~g~s^{-1}$. However, when the CME arrives at $t=24~min$, we see an inflow due to the incoming CME material pushing against the outflowing planetary material. Finally, after $90~min$, the CME penetrates through the sphere and the inflowing mass flux through it is completely associated with the CME itself. Since the planetary outflow fills the region near the planet and perhaps parts of its magnetosphere (if it  exists), it is expected that a significant amount of mass will be stripped by the CME. Estimating this amount of mass-loss requires taking into account the time scales involved in filling the magnetosphere and stripping it, as well as some more consistent escape mechanisms than the one we use here. Thus, we leave this investigation for future studies.

\subsection{The Escaping Neutral Atmosphere} \label{sec:Neutrals}

Although the escaping neutral hydrogen should not be affected by the magnetic field carried by the stellar wind, one can ask whether the neutral hydrogen atoms travel freely without interacting with the ionized stellar wind and magnetospheric plasma.

%{\bf Old para:}\\
%Using the density structure from our simulations ($n \leq 10^7$~cm$^{-3}$ see Figure~\ref{fig:CasesResults}), and a
%cross section for charge-exchange $\sim  1 \AA^2$ for 20 $km/s$ neutral hydrogen, 
%cross-section for collisions between protons and neutral hydrogen atoms of $\sim  10^{-14}$~cm$^2$ for the temperatures of interest here \citep{Hunter1977},
%we find that the mean free path for the smaller kinetic energy of the neutrals near the planet is of the order of 100~km, and so much less than $1R_p$. Thus, it is reasonable to believe that the neutral material will follow the magnetic field structure to some extent via the collisions with the ions. The inclusion of the magnetized environment (i.e., MHD approach) is then crucial for rigorous study pf the interaction of the escaping atmosphere and the space environment.

The stellar wind particle density from our simulations in the vicinity of the planet (Figure~\ref{fig:Cases_pressure}) is in the range $n \sim 10^4$-$10^5$~cm$^{-3}$. The 
cross-section for collisions between protons and neutral hydrogen atoms is dominated by the elastic collision term and is of the order $\sim  10^{-14}$~cm$^2$ for the temperatures of interest here \citep{Hunter1977}. The mean free path for the low kinetic energy neutrals near the planet between collisions with stellar wind protons is then of the order of 10,000-100,000~km, or similar to the planetary radius of $R_p=25,930$~km \citep{Martioli21}. Thus, it is reasonable to believe that the neutral material will follow the magnetic field structure to some extent via the collisions with the ions. The inclusion of the magnetized environment (i.e., MHD approach) is then crucial for rigorous study of the interaction between the escaping atmosphere and the space environment. \cite{McCann19} estimated that the timescale for ionization of the escaping neutrals is about 4 hours. With a maximum escape velocity of $20~km~s^{-1}$, these neutrals could reach a distance of about 10 planetary radii before they get ionized. However, it is more likely that their escape velocity is much less than the above upper limit, and that the neutrals will collide with other neutrals/ions much closer to the planet. All of the above suggest that the scenario of a neutral escaping atmosphere that freely expands to large distances from the planet is very unlikely, especially in a close-in orbits, where the photoevaporation is expected to be very strong. It is important to note here that instead of a pure hydrodynamic evaporation, other escape mechanisms may lead to the significant loss of the planetary gaseous envelop \citep[see recent review by][]{Gronoff2020}. 

The Ly$\alpha$ absorption we obtain using our assumption of a 50\% ionization fraction of our single-fluid plasma provides an absorption profile with 10-20\% flux reduction of the blue wing. In order to further investigate the size and shape of the escaping neutral atmosphere, we repeat our simulation for Cases 1-3 and include a neutral hydrogen fluid. The neutral fluid is imposed at the inner boundary in the same manner as the single-fluid plasma, and it expands from the inner boundary. It is important to note that in this numerical experiment, the plasma and neutral fluids are {\it completely decoupled}, and no interaction between them is included (including the collisional analysis described above). Of course, in this formalism the neutral fluid is not affected by the magnetic field and it is purely hydrodynamic. 

\begin{figure*}[h!]
\centering
\includegraphics[width=6.75in]{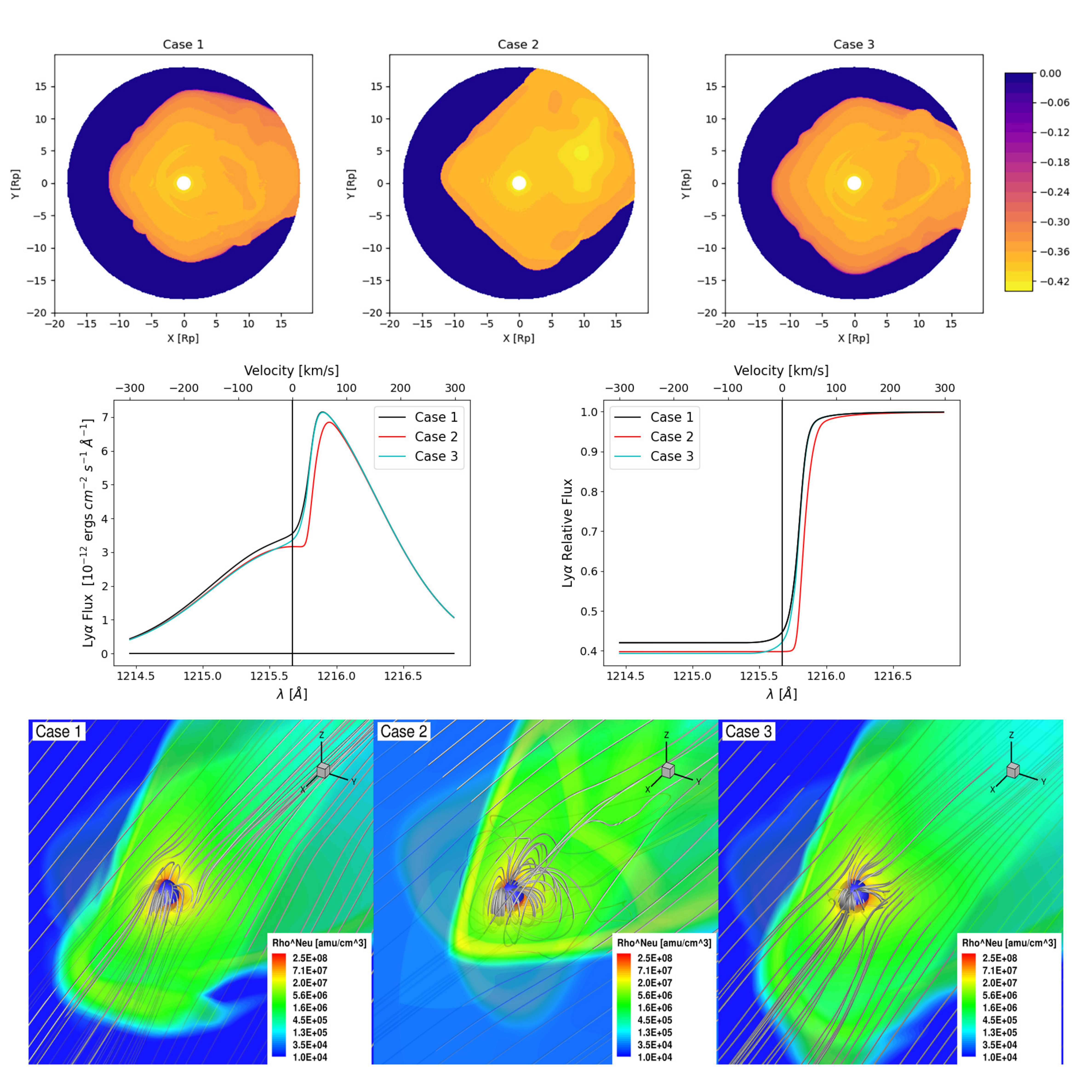}
\caption{Similar display as in Figure~\ref{fig:CasesResults} but for the neutral hydrogen fluid.}
\label{fig:CasesResultsNeutrals}
\end{figure*}

The results of our simulations for the neutral hydrogen fluid are shown in Figure~\ref{fig:CasesResultsNeutrals}. The display is the same as Figure~\ref{fig:CasesResults}. One can see that the neutrals expand significantly far from the planet, beyond 10 planetary radii as predicted by the conventional view of an extended neutral atmosphere in close-in gas planets. However, when using this extended neutral atmosphere to calculate the Ly$\alpha$ absorption profile, we find that the blue wing is almost completely absorbed (while the red wing is not). Even though the neutral gas should not be affected by the magnetic field, it is relatively coupled to the stellar wind. Thus, in the positive x direction (red-shift, away from the observer), the neutrals’ motion is stopped by the incoming stellar wind, while in the negative x direction (blue-shift, towards the observer), the neutrals are carried by the ion motion, as well as their own, slow motion. Our neutral experiment suggests that a very extended neutral atmosphere will be manifested in a greater blue-wing absorption. Of course, our simulation here is over-simplified. A better treatment would include a detailed ion-neutral interaction, self-consistent ionization, and ion escape in a multi-species manner. We plan to implement these processes in future extension of this work. 

The low escape velocities we obtain are consistent with the concept that the Ly$\alpha$ absorption profile is blue-shifted mostly due to the stellar wind - outflow interaction and not due to an extreme planetary outflow by itself \citep{Murray-Clay2009,McCann19,Carolan20}. Moreover, our MHD simulations have even stronger coupling between the SW and the planetary magnetosphere than these past hydrodynamic simulations. In our simulations, the fast SW (and even faster CME) compress near the planet, leading to a high enough density to contribute to the synthetic absorption line profile. In reality, the ionized SW may need to undergo charge-exchange in order to contribute to the Ly$\alpha$ absorption as proposed by \cite{Holmstrom2008,Murray-Clay2009,McCann19,Debrecht2022}. Such charge-exchange is seen in many bodies in our own solar system, and it is expected to be stronger in close-in exoplanets, where the SW density is orders of magnitude higher than the SW density around solar system bodies. 

The possibility that the observed Ly$\alpha$ absorption is due to the SW and not the planetary escaping atmosphere is potentially problematic for observing the planetary atmosphere to extend our knowledge about atmospheric loss and, by extension, exoplanet evolution. However, this opens the door to use these observations to constrain the SW itself. Our work also raises the possibility of a future study about how the planetary magnetic field affects the Ly$\alpha$ profile (here we only use a single planetary field value of $0.3~G$). If Ly$\alpha$ observations could provide information about the planetary magnetic field, then its contribution to exoplanet evolution may still be significant. 

\subsection{the Possible Role of the Planetary Magnetic Field Strength}
\label{sec:MagneticFieldRole}

The possible role of the planetary magnetic field on the escaping neutrals and their pattern is not clear. It is reasonable to assume that a stronger planetary field would probably reduce the sensitivity of the outflow to the stellar wind conditions since a stronger field would keep a more steady, dipolar structure close to the planet, where the outflow operates \cite[see, e.g.,][]{Owen2014,Owen19,Khodachenko2021}. 

The impact of the magnetic field on the escape itself is a completely open question. The intuitive assumption is that a stronger planetary field would protect the atmosphere from stellar wind stripping. However, at the upper atmosphere, ion chemistry dominates, and the impact of the planetary field is stronger. It has been shown that in some cases the strong field can actually enhance the escape via wave-particle interaction heating \citep{Strangeway10a,Strangeway10b,Strangeway2019}. This is a very broad subject that is beyond the scope of our paper, especially due to the fact that our model here cannot capture any of these processes. This should be studied by a model for the upper atmosphere, which includes ion chemistry, electrodynamics, and the planetary field. 

A stronger field could potentially impact the amount of neutral Hydrogen, which absorbs the Ly alpha flux. If the Ly alpha signature is produced by charge-exchanged stellar wind, then a stronger field would push the incoming stellar wind further from the planet, where the planetary neutral density is much lower. We would expect that this would lead to a reduction in the Ly alpha absorption.

\section{Summary \& Conclusions} \label{sec:conclusions}

In this paper, we simulate the response of the atmospheric outflow from AU Mic to the time-varying conditions of the stellar wind along the orbit, and to a stellar CME. We also calculated the predicted Ly$\alpha$ transit absorption signature of this interaction. We find that:
\begin{enumerate}
    \item The Ly$\alpha$ emission is highly variable due to the highly-varying stellar wind conditions (ranging between 10-20\% absorption in our simulations) through an orbit. Thus, Ly$\alpha$ observations represent a particular epoch, which may change even over a fraction of the exoplanet orbit. Thus, one epoch alone should not be used to deduce information about the global system, e.g., total escape rate and the planetary magnetic field strength. Instead, multi-epoch Ly$\alpha$ profiles could be useful for characterizing these parameters. 
    \item The pressure gradient, which drives planetary atmosphere hydrodynamic escape, may be weak in close-in exoplanets due to the relatively high pressure of the nearby space environment. During CME events, the pressure gradient may be completely removed, thereby suppressing  atmospheric escape. 
    \item Our results show a strong impact of the magnetic field and magnetic pressure on the escaping atmospheric material and the Ly$\alpha$ profile. Thus, our results show that an MHD, and not hydrodynamic treatment is crucial for the study of escaping atmospheres and their interaction with the nearby space environment. Ideally, a multi-fluid approach that couples ions and neutrals would be the most complete. 
    \item Our results support the claim that the high Doppler blue-shifted velocity is likely attributed to the stellar wind sweeping the escaping neutral material, or it might be due to charge-exchanged, neutralized fast stellar wind. It is unlikely that the atmospheric neutral material reaches such velocities.
    \item Our CME simulations indicate a potential strong stripping of magnetospheric material from the planet, including some of the neutral escaping atmosphere. However, without a consistent model for the atmospheric escape, our simulations lack the ability to properly quantify the rate of mass-loss per CME event. 
    
\end{enumerate}

\section*{acknowledgments}

We than an unkown referee for her/his useful comments. This work is supported by NASA XRP grant 80NSSC20K0840. FF was supported, in part, by  NASA through Chandra Theory Award Number $TM0-21001X$ issued by the Chandra X-ray Observatory Center, which is operated by the Smithsonian Astrophysical Observatory for and on behalf of NASA under contract NAS8-03060. LMH has received funding from the European Research Council (ERC) under the European Union’s Horizon 2020 research and
innovation program (grant agreement No. 853022, PEVAP). JJD was supported by NASA contract NAS8-39073 to the Chandra X-ray Center and thanks the Director, P. Slane,
for continuing advice and support. Simulation results were obtained using the (open source) Space Weather Modeling Framework, developed by the Center for Space Environment Modeling, at the University of Michigan with funding support from NASA ESS, NASA ESTO-CT, NSF KDI, and DoD MURI. The simulations were performed on NASA's Pleiades cluster (under SMD-20-52848317) and on the Massachusetts Green High Performance Computing Center (MGHPCC) cluster.
 
%\end{acknowledgments}

%\bibliography{AUMic}{}
%\bibliographystyle{aasjournal}

%%%%%%%%%%%%%%%%%%%%%%%%%
%  Tables
%%%%%%%%%%%%%%%%%%%%%%%%%

\begin{table*}[h!]
\begin{center}
\begin{tabular}{  p{1in}  p{2in} }
%\begin{tabular}{ccc}
\hline
{\bf Case} & {\bf Mass-loss Rate [$g~s^{-1}$]}\\
\hline
1 & $-2\cdot 10^{10}$ (outflow)\\
2 & $-2\cdot 10^{10}$ (outflow)\\
3 & $-2\cdot 10^{10}$ (outflow)\\
\hline
{\bf CME Time [m]} & {\bf Mass-loss Rate [$g~s^{-1}$]}\\
0 & $-1.8\cdot 10^{10}$ (outflow)\\
24 & $7\cdot 10^8$ (inflow)\\
90 & $3\cdot 10^9$ (inflow)\\
\hline
\end{tabular}
\end{center}
\caption {Ion and neutral mass-loss rates during the CME event.} 
\label{table1}
\end{table*}

%%%%%%%%%%%%%%%%%%%%%%%%%
%  Figures
%%%%%%%%%%%%%%%%%%%%%%%%%

\end{document}